\documentclass[12pt]{article}
\usepackage{a4}
\usepackage{amssymb}
\usepackage{cite}
\newcommand{\be}{\begin{equation}}
\newcommand{\ee}{\end{equation}}
\newcommand{\bea}{\begin{eqnarray}}
\newcommand{\eea}{\end{eqnarray}}

\usepackage{hyperref}
\begin{document}
\thispagestyle{empty}
\def\thefootnote{\fnsymbol{footnote}}
\begin{center}\Large
On mini-superspace limit of boundary three-point function in Liouville field theory \\
\vskip 2em
November 2017
\end{center}\vskip 0.2cm
\begin{center}
Elena Apresyan$^{1}$
\footnote{elena-apresyan@mail.ru}
 and Gor Sarkissian$^{1,2}$\footnote{ gor.sarkissian@ysu.am}
\end{center}
\vskip 0.2cm
\begin{center}
$^1$ Yerevan Physics Institute, \\
Alikhanian Br. 2, 0036\, Yerevan, Armenia\\
\end{center}
\begin{center}
$^2$Department of Physics, \ Yerevan State University,\\
Alex Manoogian 1, 0025\, Yerevan, Armenia\\
\end{center}

\vskip 1.5em
\begin{abstract} \noindent
We study mini-superspace semiclassical limit of the boundary three-point function in the Liouville field theory.
We compute also matrix elements for the Morse potential quantum mechanics.
An exact agreement between the former and the latter is found.
 We show that both of them are given by the generalized hypergeometric functions.

\end{abstract}


\newpage
\tableofcontents
\newpage
\section{Introduction}
Recently the various semiclassical limits of the Liouville correlation functions appeared in different instances.
For example we can  mention study of conformal blocks in AdS/CFT correspondence, see {\it e.g.} \cite{Hijano:2015qja,Alkalaev:2015wia,Fitzpatrick:2015zha}, semiclassical limits of the Nekrasov partition functions, see {\it e.g} \cite{Mironov:2009qn,Fateev:2011qa,Hama:2013ama,Piatek:2013ifa,Poghosyan:2017qdv,Poghosyan:2016lya},
minisuperspace limit of correlation functions in $\rm{AdS}_3/H_3^{+}$ \cite{Ribault:2007td,Ribault:2009ui}, semiclassical limit
of correlation functions in the presence of defects and boundaries \cite{Fateev:2010za,Poghosyan:2015oua}
 and the most recently found application
of the semiclassical limit of Liouville field theory to the SYK problem \cite{Mertens:2017mtv}.

In this paper we study matrix elements of the boundary Liouville field theory in  mini-superspace limit. In the mini-superspace limit
one considers a limit where only the zero mode dynamics survives and the theory is reduced to the corresponding quantum mechanical problem.
 The mini-superspace limit of the Liouville field theory was considered
in \cite{Braaten:1982yn,Braaten:1983np}. In these papers the matrix elements of the Liouville quantum mechanics with the exponential potential
were computed.
Later it was shown in \cite{Thorn:2002am} that the DOZZ structure constants \cite{Dorn:1994xn,Zamolodchikov:1995aa} in this limit coincide with the matrix elements
found in \cite{Braaten:1982yn,Braaten:1983np}. It was also demonstrated in \cite{Zamolodchikov:1995aa} that the Liouville two-point function
in the mini-superspace limit in agreement with the reflection function of the Liouville quantum mechanics eigenfunctions given by
the modified Bessel function.
In papers \cite{Bajnok:2007ep,Dorn:2008sw} was studied the mini-super space limit of the boundary Liouville field theory (BLFT).
It was found that BLFT in this limit reduced to the Morse potential quantum mechanics. It was shown in \cite{Bajnok:2007ep} that in the mini-super space
limit the boundary two-point function, computed in \cite{Fateev:2000ik}, coincides with the reflection amplitude of the eigenfunctions of the Morse potential Hamiltonian given by the Whittaker functions.

In this paper we study the mini-superspace limit of the boundary three-point function in the BLFT.
The boundary three-point function in the BLFT was computed in \cite{Ponsot:2001ng} and expressed vie double Gamma and double Sine functions \cite{gam1,gam2}.
Using the asymptotic properties of the double Gamma and Sine functions \cite{Ribault:2007td} we have shown that in the mini-superspace limit the boundary three-point function
can be expressed via the Meijer functions $G^{3,2}_{3,3}$ with the unit argument or equivalently via the generalized hypergeometric functions
${}_3F_2$ with the unit argument.
We also computed matrix elements for the Morse potential and have shown that they can be expressed via the generalized hypergeometric functions ${}_3F_2$ with the unit argument as well. Using the identities, relating different generalized hypergeometric functions with the unit argument \cite{slater,wilfrid,whipple}, and matching quantum and classical parameters, we established exact agreement
between the mini-superspace limit of the boundary three-point function and the matrix elements for the Morse potential.
It is important to note that in the BLFT relation of the boundary cosmological parameter to the corresponding quantum parameter
appearing in the boundary one-point function is twofold due to a sign ambiguity in the choice of the square root branch.
We found that to match the minisuperspace limit of the boundary three-point  with the corresponding quantum mechanical matrix element
we should use the branch with the negative sign. We also found that passing from one branch to another brings to additional factor
in the normalization of the wave functions corresponding to the boundary condition changing operators. We would like also
to mention that various consequences of the branching of the BLFT parameters earlier were considered in  \cite{Teschner:2003qk}.

The paper is organized as follows. In section 2 we review the BLFT and compute the mini-superspace limit of the boundary three-point function. In section 3 we compute
matrix elements for the Morse potential and establish precise agreement with the boundary three-point function in the mini-superspace limit found in the previous section. In appendices A, B and C we review various properties of the special functions used in the paper.

\section{Boundary Liouville field theory}

Let us consider the Liouville field theory on a strip $\mathbb{R}\times [0,\pi]$ , parameterized by the time $\tau$ and space
$\sigma$ coordinates, $0\leq \sigma \leq \pi$. The conformal invariant action has the form:
\be
S=\int_{-\infty}^{\infty}d\tau\int_0^{\pi}d\sigma \left({1\over 4\pi}(\partial_a\phi)^2+\mu e^{2b\phi}\right)+
\int_{-\infty}^{\infty}d\tau M_1 e^{b\phi}|_{\sigma=0}+\int_{-\infty}^{\infty}d\tau M_2 e^{b\phi}|_{\sigma=\pi}
\ee
where $M_1$ and $M_2$ are the corresponding boundary cosmological constants.

Let us review some facts on the boundary Liouville field theory \cite{Fateev:2000ik,Teschner:2000md,Ponsot:2001ng}.
The primary fields of the Liouville field theory are $V_{\alpha}$, associated with the vertex operators $e^{2\alpha \phi}$.
They have conformal dimension
\be
\Delta_{\alpha}=\alpha(Q-\alpha),\quad Q=b+{1\over b}
\ee

In the presence of the boundary with the cosmological constant $M$ the primary fields $V_{\alpha}$ have the one-point functions:
\be
\langle 0|V_{\alpha}(z,\bar{z})|0\rangle ={U_{\sigma}(\alpha)\over |z-\bar{z}|^{2\Delta_{\alpha}}}
\ee
where
 \be
U_{\sigma}(\alpha)={2\over  b}(\pi\mu\gamma(b^2))^{(Q-2\alpha)/2b}\Gamma(1-b(Q-2\alpha))\Gamma(-b^{-1}(Q-2\alpha))
\cos(\pi (2\sigma-Q)(2\alpha-Q))
\ee
where the parameter $\sigma$ is related to the  boundary cosmological constant $M$ by the relation:
\be\label{bnpar}
M=\sqrt{\mu\over \sin(\pi b^2)}\cos \pi b\left(2\sigma-Q\right)
\ee

Besides bulk primary fields in the boundary conformal field theory exist also boundary condition changing operators, parameterized
by the types of the switched boundary conditions and conformal weight. In the case of the BLFT they are given by the fields
$\Psi_{\beta}^{\sigma_1\sigma_2}$ with conformal weight $\Delta_{\beta}=\beta(\beta-Q)$. They have the two-point function:
\be
\langle 0|\Psi_{\beta_1}^{\sigma_1\sigma_2}(x)\Psi_{\beta_2}^{\sigma_2\sigma_1}(0)|0\rangle ={
\delta(\beta_2+\beta_1-Q)+S(\beta_1,\sigma_2,\sigma_1)\delta(\beta_2-\beta_1)\over |x|^{2\Delta_{\beta_1}}}
\ee
where
\bea\label{twobo}
&&S(\beta,\sigma_2,\sigma_1)=\left(\pi\mu\gamma(b^2)b^{2-2b^2}\right)^{Q-2\beta\over 2b}\times\\ \nonumber
&&\times {\Gamma_b(2\beta-Q)\over \Gamma_b(Q-2\beta)}{S_b(\sigma_2+\sigma_1-\beta)S_b(2Q-\sigma_2-\sigma_1-\beta)\over
S_b(\sigma_2-\sigma_1+\beta)S_b(\sigma_1-\sigma_2+\beta)}
\eea

and three-point function

\bea
\langle 0|\Psi^{\sigma_1\sigma_3}_{\beta_3}(x_3)\Psi^{\sigma_3\sigma_2}_{\beta_2}(x_3)\Psi^{\sigma_2\sigma_1}_{\beta_1}(x_3)|0\rangle=\\ \nonumber
{C^{\sigma_3\sigma_2\sigma_1}_{\beta_3\beta_2\beta_1}\over
|x_{21}|^{\Delta_1+\Delta_2-\Delta_3}|x_{32}|^{\Delta_2+\Delta_3-\Delta_1}|x_{31}|^{\Delta_3+\Delta_1-\Delta_2}}
\eea

\be
C^{\sigma_3\sigma_2\sigma_1}_{\beta_3|\beta_2\beta_1}\equiv C^{\sigma_3\sigma_2\sigma_1}_{Q-\beta_3,\beta_2,\beta_1}
\ee

\be
C^{\sigma_3\sigma_2\sigma_1}_{\beta_3|\beta_2\beta_1}=R_{\sigma_2,\beta_3}\left[\begin{array}{cc}
\beta_2& \beta_1\\
\sigma_3& \sigma_1 \end{array}\right]\int_{-i\infty}^{i\infty}{d\tau\over i} J_{\sigma_2,\beta_3}\left[\begin{array}{cc}
\beta_2& \beta_1\\
\sigma_3& \sigma_1 \end{array}\right]
\ee
where

\bea\label{rsibe}
&&R_{\sigma_2,\beta_3}\left[\begin{array}{cc}
\beta_2& \beta_1\\
\sigma_3& \sigma_1 \end{array}\right]=(\pi\mu\gamma(b^2)b^{2-2b^2})^{{1\over 2b}(\beta_3-\beta_2-\beta_1)}
\\ \nonumber
&\times&{\Gamma_b(2Q-\beta_1-\beta_2-\beta_3)\Gamma_b(\beta_2+\beta_3-\beta_1)\Gamma_b(Q+\beta_2-\beta_1-\beta_3)\Gamma_b(Q+\beta_3-\beta_2-\beta_1)\over
\Gamma_b(2\beta_3-Q)\Gamma_b(Q-2\beta_2)\Gamma_b(Q-2\beta_1)\Gamma_b(Q)}
\\ \nonumber
&\times& {S_b(\beta_3+\sigma_1-\sigma_3)S_b(Q+\beta_3-\sigma_3-\sigma_1)\over S_b(\beta_2+\sigma_2-\sigma_3)S_b(Q+\beta_2-\sigma_3-\sigma_2)}
\eea
and
\bea\label{int123}
&&J_{\sigma_2,\beta_3}\left[\begin{array}{cc}
\beta_2& \beta_1\\
\sigma_3& \sigma_1 \end{array}\right]=
 {S_b(U_1+\tau)S_b(U_2+\tau)\over
S_b(V_1+\tau)S_b(V_2+\tau)}
{S_b(U_3+\tau)S_b(U_4+\tau)\over
S_b(V_3+\tau)S_b(V_4+\tau)}
\eea
\bea
&&U_1=\sigma_2+\sigma_1-\beta_1,\hspace{1.8cm} V_1=Q+\sigma_2+\beta_3-\beta_1-\sigma_3\\ \nonumber
&&U_2=Q+\sigma_2-\beta_1-\sigma_1,\hspace{1cm} V_2=2Q+\sigma_2-\beta_3-\sigma_3-\beta_1\\ \nonumber
&&U_3=\sigma_2+\beta_2-\sigma_3,\hspace{1.8cm} V_3=2\sigma_2\\ \nonumber
&&U_4=Q+\sigma_2-\beta_2-\sigma_3,\hspace{1cm} V_4=Q
\eea
$\Gamma_b(x)$ and $S_b(x)$ in the formulae above denote the double Gamma and Sine functions reviewed in appendix A.

The three-point function has the property, that setting one of the field to vacuum, one recovers the two-point function. For
example it was checked in \cite{Ponsot:2001ng} that
\be
{\rm lim}_{\beta_1 \to 0}C^{\sigma_3\sigma_2\sigma_1}_{\beta_3|\beta_2\beta_1}=\delta(\beta_3-\beta_2)+
S(\beta_2,\sigma_3,\sigma_2)\delta(\beta_3+\beta_2-Q)
\ee

Let us now consider the minisuperspace limit of three-point function.

As the warm-up exercise we review the minisuperspace limit of two-point function (\ref{twobo}), computed in \cite{Bajnok:2007ep}.
It is argued in \cite{Bajnok:2007ep} that one should take the limit $b\to 0$ and scale the parameters $\beta$ and $\sigma$ in the following way:
\be\label{beta1}
\beta={Q\over 2}+ikb
\ee
and
\bea\label{sigma12}
&&\sigma_1={1\over 4b}+\rho_1 b\\ \nonumber
&&\sigma_2={1\over 4b}+\rho_2 b
\eea

Using formulae (\ref{s3}, (\ref{s4}) and (\ref{g1}) in appendix A one can easily obtain:
\be\label{tupson}
 S(\beta,\sigma_2,\sigma_1)\to \left({4\pi\mu\over b^2}\right)^{-ik}{\Gamma(2ik)\over \Gamma(-2ik)}{\Gamma\left(\rho_1+\rho_2-{1\over 2}-ik\right)\over
\Gamma\left(\rho_1+\rho_2-{1\over 2}+ik\right)}
\ee

To compute the mini-superspace limit of the boundary three-point function  we will use  the ansatz (\ref{sigma12}) for all the three  boundary condition parameters:
\bea\label{sigma123}
&&\sigma_1={1\over 4b}+\rho_1 b\\ \nonumber
&&\sigma_2={1\over 4b}+\rho_2 b\\ \nonumber
&&\sigma_3={1\over 4b}+\rho_3 b
\eea

For the primary fields parameters we will use the ansatz suggested in \cite{Thorn:2002am} for calculation of the mini-superspace limit
of the bulk three-point function:
\bea\label{beta123}
&&\beta_1={Q\over 2}+ik_1b\\ \nonumber
&&\beta_2=\eta b\\ \nonumber
&&\beta_3={Q\over 2}+ik_2b
\eea
It is convenient to denote
\be
\rho_1+\rho_2=1-\lambda
\ee
\be\label{xi23}
\rho_2-\rho_3=\xi
\ee
implying also
\be
\rho_1+\rho_3=1-\lambda-\xi
\ee

Inserting (\ref{sigma123}) and (\ref{beta123}) in (\ref{int123}), using the formulas (\ref{s3}), (\ref{s4}),(\ref{s5}) in appendix A,
and rescaling the integration variable $\tau\to b\tau$, one obtains in the limit $b\to 0$

\bea
&&\int_{-i\infty}^{i\infty}{d\tau\over i} J_{\sigma_2,\beta_3}\left[\begin{array}{cc}
\beta_2& \beta_1\\
\sigma_3& \sigma_1 \end{array}\right]\to
 2^{-7/2}(\pi b^2)^{-\lambda+ik_1}b^{-1}\pi^{-2}\times\\ \nonumber
&&\int_{-i\infty}^{i\infty}{d\tau\over i} {\Gamma(-\tau)\Gamma(\tau-ik_1+1/2-\lambda)\Gamma(\eta+\xi+\tau)\Gamma(ik_1-ik_2-\xi-\tau)\Gamma(ik_2+ik_1-\xi-\tau)\over \Gamma(\eta-\xi-\tau)}
\eea

Using the definition of the Meijer G-functions, reviewed in appendix B, one can write
\bea\label{g233}
&&\int_{-i\infty}^{i\infty}{d\tau\over i} J_{\sigma_2,\beta_3}\left[\begin{array}{cc}
\beta_2& \beta_1\\
\sigma_3& \sigma_1 \end{array}\right]\to\\ \nonumber
 && 2^{-5/2}(\pi b^2)^{-\lambda+ik_1}b^{-1}\pi^{-1}G^{3,2}_{3,3}
 \left(1\Bigg|\begin{array}{c}
{1\over 2}+\lambda+ik_1,1-\eta-\xi,\eta-\xi\\
0, ik_1-ik_2-\xi,ik_1+ik_2-\xi\end{array}\right)=\quad\quad\\ \nonumber
&& 2^{-5/2}(\pi b^2)^{-\lambda+ik_1}b^{-1}\pi^{-1}G^{3,2}_{3,3}
 \left(1\Bigg|\begin{array}{c}
{1\over 2}+\lambda+\xi+ik_1,1-\eta,\eta\\
\xi, ik_1-ik_2,ik_1+ik_2\end{array}\right)
\eea
In the second line we used the identity (\ref{gxi}) in appendix B.

For further purposes, it is convenient to present the Meijer $G^{3,2}_{3,3}$-function (\ref{g233}) in a special way, use of which become clear
in the next section. Namely, first we decompose the  $G^{3,2}_{3,3}$-function as a sum of ${}_3F_2$ hypergeometric functions with the unit argument
according to eq. (\ref{g3332}) in appendix B. Afterwards we transform obtained in this way ${}_3F_2$ hypergeometric functions with the unit argument
successively applying identities (\ref{f321}) and (\ref{f322}) in appendix C. We end up with
\bea\label{glim}
&&G^{3,2}_{3,3}
 \left(1\Bigg|\begin{array}{c}
{1\over 2}+\lambda+ik_1+\xi,1-\eta,\eta\\
\xi, ik_1-ik_2,ik_1+ik_2\end{array}\right)={\Gamma(\xi+\eta)\Gamma({1\over 2}+\lambda-ik_1)\over \sin\pi(ik_1+{1\over 2}+\lambda)}\times\quad\\ \nonumber
&&\left[{\Gamma(2ik_2)\Gamma(ik_1-ik_2+\eta)\Gamma({1\over 2}-ik_2-\lambda-\xi)\over \Gamma(-ik_1+ik_2+\eta)\Gamma(-ik_2+{1\over 2}+\lambda+\eta)\Gamma(-ik_2+{1\over 2}-\lambda+\eta)\Gamma(ik_2+{1\over 2}+\lambda-\eta)}\right.\times\nonumber\\
&&{}_3F_2\left(\begin{array}{c}
-ik_1-ik_2+\eta, ik_1-ik_2+\eta,{1\over 2}+\lambda+\xi-ik_2; \\ \nonumber
1-2ik_2, {1\over 2}+\lambda-ik_2+\eta: 1\end{array}\right)+\nonumber \\
&&{\Gamma(-2ik_2)\Gamma(ik_1+ik_2+\eta)\Gamma({1\over 2}+ik_2-\lambda-\xi)\over \Gamma(-ik_1-ik_2+\eta)\Gamma(ik_2+{1\over 2}+\lambda+\eta)\Gamma(ik_2+{1\over 2}-\lambda+\eta)\Gamma(-ik_2+{1\over 2}+\lambda-\eta)}\times\nonumber\\
&&\left.{}_3F_2\left(\begin{array}{c}
ik_1+ik_2+\eta, -ik_1+ik_2+\eta,{1\over 2}+\lambda+\xi+ik_2; \\ \nonumber
1+2ik_2, {1\over 2}+\lambda+ik_2+\eta: 1\end{array}\right)\right]
\eea

Now inserting (\ref{sigma123}) and (\ref{beta123}) in (\ref{rsibe}), and using formulae (\ref{s1})-(\ref{g2}) in appendix A,
we obtain for the prefactor (\ref{rsibe}) in the limit $b\to 0$
\bea\label{rlim}
&&R_{\sigma_2,\beta_3}\left[\begin{array}{cc}
\beta_2& \beta_1\\
\sigma_3& \sigma_1 \end{array}\right]\to \left({4\pi\mu \over b^{2}}\right)^{(ik_2-ik_1-\eta)/2}4(\pi b^2)^{-ik_1+\lambda}b\pi^{3/2}\\ \nonumber
&&{\Gamma(-ik_1+ik_2+\eta)\Gamma(-ik_1-ik_2+\eta)\over \Gamma(2ik_2)\Gamma(-2ik_1)\Gamma({1\over 2}-ik_2-\lambda-\xi)\Gamma(\eta+\xi)}
\eea
Combining (\ref{rlim}) and (\ref{glim}) finally we obtain\footnote{probably up to some inessential numerical factors} :
\bea\label{csigma23}
&&C^{\sigma_3\sigma_2\sigma_1}_{\beta_3|\beta_2\beta_1}\to C^{\lambda\xi}_{k_2|\eta k_1}=\\ \nonumber
&&\left({4\pi\mu \over b^{2}}\right)^{(ik_2-ik_1-\eta)/2}
{\Gamma({1\over 2}+\lambda-ik_1)\over \sin\pi(ik_1+{1\over 2}+\lambda)\Gamma(2ik_2)\Gamma(-2ik_1)\Gamma({1\over 2}-ik_2-\lambda-\xi)}\times\quad\quad\\ \nonumber
&&\left[{\Gamma(2ik_2)\Gamma(ik_1-ik_2+\eta)\Gamma(-ik_1-ik_2+\eta)\Gamma({1\over 2}-ik_2-\lambda-\xi)\over \Gamma(-ik_2+{1\over 2}+\lambda+\eta)\Gamma(-ik_2+{1\over 2}-\lambda+\eta)\Gamma(ik_2+{1\over 2}+\lambda-\eta)}\right.\times\nonumber\\
&&{}_3F_2\left(\begin{array}{c}
-ik_1-ik_2+\eta, ik_1-ik_2+\eta,{1\over 2}+\lambda+\xi-ik_2; \\ \nonumber
1-2ik_2, {1\over 2}+\lambda-ik_2+\eta: 1\end{array}\right)+\nonumber \\
&&{\Gamma(-2ik_2)\Gamma(ik_1+ik_2+\eta)\Gamma(-ik_1+ik_2+\eta)\Gamma({1\over 2}+ik_2-\lambda-\xi)\over \Gamma(ik_2+{1\over 2}+\lambda+\eta)\Gamma(ik_2+{1\over 2}-\lambda+\eta)\Gamma(-ik_2+{1\over 2}+\lambda-\eta)}\times\nonumber\\
&&\left.{}_3F_2\left(\begin{array}{c}
ik_1+ik_2+\eta, -ik_1+ik_2+\eta,{1\over 2}+\lambda+\xi+ik_2; \\ \nonumber
1+2ik_2, {1\over 2}+\lambda+ik_2+\eta: 1\end{array}\right)\right]
\eea
As we will see in the next section, especially  important role plays the case when  $\xi=-\eta$.
For $\xi=-\eta$ (\ref{csigma23}) simplifies and takes the form:
\bea\label{csigma234}
&& C^{\lambda(-\eta)}_{k_2|\eta k_1}=\\ \nonumber
&&\left({4\pi\mu \over b^{2}}\right)^{(ik_2-ik_1-\eta)/2}
{\Gamma({1\over 2}+\lambda-ik_1)\over \sin\pi(ik_1+{1\over 2}+\lambda)\Gamma(2ik_2)\Gamma(-2ik_1)\Gamma({1\over 2}-ik_2-\lambda+\eta)}\times\quad\quad\\ \nonumber
&&\left[{\Gamma(2ik_2)\Gamma(ik_1-ik_2+\eta)\Gamma(-ik_1-ik_2+\eta)\over \Gamma(-ik_2+{1\over 2}+\lambda+\eta)\Gamma(ik_2+{1\over 2}+\lambda-\eta)}\right.\times\nonumber\\
&&{}_3F_2\left(\begin{array}{c}
-ik_1-ik_2+\eta, ik_1-ik_2+\eta,{1\over 2}+\lambda-\eta-ik_2; \\ \nonumber
1-2ik_2, {1\over 2}+\lambda-ik_2+\eta: 1\end{array}\right)+\nonumber \\
&&{\Gamma(-2ik_2)\Gamma(ik_1+ik_2+\eta)\Gamma(-ik_1+ik_2+\eta)\over \Gamma(ik_2+{1\over 2}+\lambda+\eta)\Gamma(-ik_2+{1\over 2}+\lambda-\eta)}\times\nonumber\\
&&\left.{}_3F_2\left(\begin{array}{c}
ik_1+ik_2+\eta, -ik_1+ik_2+\eta,{1\over 2}+\lambda-\eta+ik_2; \\ \nonumber
1+2ik_2, {1\over 2}+\lambda+ik_2+\eta: 1\end{array}\right)\right]
\eea

Let us consider the limit $\beta_2\to 0$ and correspondingly $\eta\to 0$.

Using, that as we explained in appendix C, in this limit ${}_3F_2$  reduces to  ${}_2F_1$, which for the unit argument is given by eq. (\ref{f12}),
it is straightforward to show that:
\be\label{tup}
{\rm lim}_{\eta \to 0}C^{\lambda(-\eta)}_{k_2|\eta k_1}=\delta(k_1-k_2)+
\left({4\pi\mu\over b^2}\right)^{-ik_1}{\Gamma(2ik_1)\over \Gamma(-2ik_1)}{\Gamma\left({1\over 2}-\lambda-ik_1\right)\over
\Gamma\left({1\over 2}-\lambda+ik_1\right)}\delta(k_1+k_2)
\ee
in agreement with (\ref{tupson}).

\section{Matrix elements in the Morse potential}

In the mini-superspace limit the boundary Liouville field theory is described by the Hamiltonian with the Morse potential \cite{Bajnok:2007ep,Dorn:2008sw}.
The corresponding eigenfuntions satisfy the Schr\"odinger equation:

\be\label{schrod}
-{\partial^2\psi\over \partial\phi_0^2}+\pi\mu e^{2b\phi_0}\psi+(M_1+M_2)e^{b\phi_0}\psi=k^2b^2\psi
\ee

The relation between parameters $M_i$ appearing in the Schr\"odinger equation and parameters $\rho_i$ used in the previous section
can be found using (\ref{sigma123}) and (\ref{bnpar}) and taking the limit $b\to 0$:
\be
M_i=\sqrt{\mu\over \sin(\pi b^2)}\sin \pi b^2 (2\rho_i-1)\to \pm(\mu\pi)^{1/2}b(2\rho_i-1)
\ee
The solution of the eq. (\ref{schrod}) is given by the Whittaker functions $W_{\mu,\nu}(y)$ \cite{ryzhik,baterd}:

\bea
&&\psi={\cal N}\left[e^{-y/2}y^{ik}{\Gamma\left(-2ik\right)\over \Gamma\left({1\over 2}-ik+{M_1+M_2\over 2b\sqrt{\pi\mu}}\right)}
 {}_1F_1\left({1\over 2}+ik+{M_1+M_2\over 2b\sqrt{\pi\mu}}, 1+2ik,y\right)
+\right.\nonumber \\
&&\left.e^{-y/2}y^{-ik}{\Gamma\left(2ik\right)\over \Gamma\left({1\over 2}+ik+{M_1+M_2\over 2b\sqrt{\pi\mu}}\right)}
{}_1F_1\left({1\over 2}-ik+{M_1+M_2\over 2b\sqrt{\pi\mu}}, 1-2ik,y\right)\right]\equiv\nonumber \\
&&{\cal N}\,W_{-{M_1+M_2\over 2b\sqrt{\pi\mu}}, ik}(y)y^{-{1\over 2}}
\eea
where
\be
y={2\sqrt{\pi\mu}\over b}e^{b\phi_0}
\ee
${\cal N}$ is the normalization and ${}_1F_1(a,c,z)$ is the confluent hypergeometric function:
\be
{}_1F_1(a,c,z)={\Gamma(c)\over \Gamma(a)}\sum_{n=0}^{\infty}{\Gamma(a+n)\over \Gamma(c+n)}{z^n\over n!}
\ee
Now we wish to compute matrix element of the  ``vertex operator" $e^{\eta b\phi_0}$, between the wave functions corresponding
to the boundary condition changing operators. According to this solution to the operator $\Psi^{\sigma_2\sigma_1}_{\beta_1}$ corresponds the wave function ${\cal N}_1W_{\chi_1, ik_1}(y)y^{-{1\over 2}}$ with

\be\label{chii1}
\chi_1=-{M_1+M_2\over 2b\sqrt{\pi\mu}}=\pm\lambda
\ee
and to $\Psi^{\sigma_1\sigma_3}_{\beta_3}$ corresponds the wave function ${\cal N}_2W_{\chi_2, ik_2}(y)y^{-{1\over 2}}$ with
\be\label{chii2}
\chi_2=-{M_1+M_3\over 2b\sqrt{\pi\mu}}=\pm(\lambda+\xi)
\ee
The corresponding integral can be found in \cite{ryzhik,baterd}:
\bea\label{wwnn}
&&{\cal M}_{\eta k_1k_2}^{\chi_1\chi_2}={\cal N}_1{\cal N}_2^*\int_{-\infty}^{\infty} W_{\chi_1, ik_1}(y)y^{-{1\over 2}}W_{\chi_2, -ik_2}(y)y^{-{1\over 2}}e^{\eta b\phi_0}d\phi_0=\\ \nonumber
&&={{\cal N}_1{\cal N}_2^*\over b}\left({4\pi\mu\over b^2}\right)^{-\eta/2}\int_{0}^{\infty} W_{\chi_1, ik_1}(y)W_{\chi_2, -ik_2}(y)y^{\eta-2}dy= {\cal N}_1{\cal N}_2^*(4\pi\mu b^{-2})^{-\eta/2}b^{-1}\times\\ \nonumber
&&\left[{\Gamma(ik_1-ik_2+\eta)\Gamma(-ik_1-ik_2+\eta)\Gamma(2ik_2)\over \Gamma({1\over 2}-\chi_2+ik_2)\Gamma({1\over 2}-\chi_1-ik_2+\eta)}\times\right.\\ \nonumber
&&{}_3F_2\left(\begin{array}{c}
-ik_1-ik_2+\eta, ik_1-ik_2+\eta,{1\over 2}-\chi_2-ik_2; \\
1-2ik_2, {1\over 2}-\chi_1-ik_2+\eta: 1\end{array}\right)+\\ \nonumber
&&{\Gamma(ik_1+ik_2+\eta)\Gamma(-ik_1+ik_2+\eta)\Gamma(-2ik_2)\over \Gamma({1\over 2}-\chi_2-ik_2)\Gamma({1\over 2}-\chi_1+ik_2+\eta)}\times \\ \nonumber
&&\left.{}_3F_2\left(\begin{array}{c}
ik_1+ik_2+\eta, -ik_1+ik_2+\eta,{1\over 2}-\chi_2+ik_2; \\
1+2ik_2, {1\over 2}-\chi_1+ik_2+\eta: 1\end{array}\right)\right]
\eea

Comparing (\ref{wwnn}) with (\ref{csigma234}) we see that they coincide if we set:
\be\label{ch1}
\chi_1=-\lambda
\ee
\be\label{ch2}
\chi_2=-\lambda+\eta
\ee
\be\label{caln1}
{\cal N}_1={(4\pi\mu b^{-2})^{-ik_1/2}b^{1/2}\over \sin\pi\left({1\over 2}+ik_1+\lambda\right)}{\Gamma\left({1\over 2}+\lambda-ik_1\right)\over \Gamma(-2ik_1)}
\ee
\be\label{caln2}
{\cal N}_2={1\over \pi}(4\pi\mu b^{-2})^{-ik_2/2}b^{1/2}\sin\pi\left({1\over 2}+ik_2-\lambda+\eta\right){\Gamma\left({1\over 2}+\lambda-\eta-ik_2\right)\over \Gamma(-2ik_2)}
\ee
This result leads us to the following conclusion on a role of the exponential operator $e^{\eta b\phi_0}$.
Combining (\ref{chii1}) and (\ref{chii2}) with lower signes, as indicating in (\ref{ch1}) and (\ref{ch2}), and also remembering
(\ref{sigma123}) and (\ref{xi23}) one has
\be\label{mm32}
{M_3-M_2\over 2\sqrt{\pi\mu}}=b\xi=-b\eta=\sigma_2-\sigma_3
\ee
Therefore recalling also that the exponential operator $e^{\eta b\phi_0}$ should correspond to a boundary condition changing operator
$\Psi^{\sigma_3\sigma_2}_{\beta_2}$, this result implies that the operator $e^{\eta b\phi_0}$ in the semiclassical limit produces change of the boundary condition
given by (\ref{mm32}).

It is instructive to compare the normalization of the wave functions found here with those used in \cite{Bajnok:2007ep}.
For this purpose let us compute the matrix element (\ref{wwnn}) for $\eta\to 0$ and $\chi_1=\chi_2$.
In this limit we obtain:
\bea\label{m0k}
&&{\cal M}_{0 k_1k_2}^{\chi_1\chi_1}={{\cal N}_1{\cal N}_2^*b^{-1}\Gamma(2ik_1)\Gamma(-2ik_1)
\over \Gamma({1\over 2}-\chi_1+ik_1)\Gamma({1\over 2}-\chi_1-ik_1)}\delta(k_1-k_2)+\\ \nonumber
&&{{\cal N}_1{\cal N}_2^*b^{-1}\Gamma(2ik_1)\Gamma(-2ik_1)
\over \Gamma({1\over 2}-\chi_1-ik_1)\Gamma({1\over 2}-\chi_1+ik_1)}\delta(k_1+k_2)
\eea

For $\chi_1$,  $\chi_2$, ${\cal N}_1$, ${\cal N}_2$, chosen as in (\ref{ch1})-(\ref{caln2}), with $\eta=0$, expression (\ref{m0k}) surely
coincides with the two-point function  (\ref{tup}).
But note that for
\be\label{ch11}
\chi_1=\lambda
\ee
\be\label{ch22}
\chi_2=\lambda
\ee
\be\label{caln11}
{\cal N}_1=(4\pi\mu b^{-2})^{-ik_1/2}b^{1/2}{\Gamma\left({1\over 2}-\lambda-ik_1\right)\over \Gamma(-2ik_1)}
\ee
\be\label{caln22}
{\cal N}_2=(4\pi\mu b^{-2})^{-ik_2/2}b^{1/2}{\Gamma\left({1\over 2}-\lambda-ik_2\right)\over \Gamma(-2ik_2)}
\ee
expression (\ref{m0k}) again
coincides with the two-point function  (\ref{tup}). This was established in \cite{Bajnok:2007ep}.

This shows that passing from the one branch of the square root to another introduces additional sine factors in the normalization of the
wave functions in a way to keep unchanged the two-point functions.

\section{Conclusion}

We discussed in this paper semiclassical properties of the boundary three-point functions. We found perfect agreement
with the corresponding quantum mechanical calculations. The matching of the calculations required to consider the negative branch
in the branched correspondence of the classical and quantum parameters. We  show that passing from one branch to another leads to the change in the normalization of the wave functions.
We also found the flip of the boundary conditions  induced by the exponential operators in the minisuperspace limit.

\newpage

\section*{Acknowledgments}
This work was partially carried out while the second author G.S. was visiting the high energy section of the Abdus Salam ICTP, Trieste
as a regular associate member.
We  thank  George Jorjadze for many valuable discussions.
We would like to give our special thanks to Sylvain Ribault for sharing with us his knowledge on
the asymptotic behaviour of the double Gamma and Sine functions and sending his private notes.
The work of both authors was partially supported by the Armenian SCS  grant 15T-1C308.
\\


\appendix

\section{Double Gamma and double Sine functions}
Here we review double Gamma $\Gamma_b(x)$ and double Sine $S_b(x)$ functions \cite{gam1,gam2}.

$\Gamma_b(x)$ can be defined by means of the integral representation
\bea
\log \Gamma_b(x)=\int_0^{\infty}{dt\over t}\Bigg[{e^{-xt}-
e^{-Qt/2}\over (1-e^{-bt})(1-e^{-t/b})}-
{(Q-2x)^2\over 8e^t}-{Q-2x\over t}\Bigg]\, .
\eea
It has the property:
\be
\Gamma_b(x+b)=\sqrt{2\pi}b^{bx-{1\over 2}}\Gamma^{-1}(bx)\Gamma_b(x)
\ee
The double Sine function $S_b(x)$ may be defined in term of $\Gamma_b(x)$ as

\be\label{s1}
S_b(x)={\Gamma_b(x)\over \Gamma_b(Q-x)}\, .
\ee
It has an integral representation:
\be
\log S_b(x)=\int_0^{\infty} {dt\over t}\left({\sinh t(Q-2x)\over 2\sinh bt\sinh b^{-1}t}-{Q-2x\over 2t}\right)\, .
\ee
and the properties:
\be\label{s2}
S_b(x+b)=2\sin(\pi bx)S_b(x)
\ee
\be
S_b(x+1/b)=2\sin(\pi x/b)S_b(x)
\ee
For $b\to 0$ the double Gamma $\Gamma_b(x)$ and double Sine $S_b(x)$ functions have the asymptotic behaviour \cite{Ribault:2007td}:
\be\label{s3}
S_b(bx)\to (2\pi b^2)^{x-{1\over 2}}\Gamma(x)
\ee
\be\label{s4}
S_b\left({1\over 2b}+bx\right)\to 2^{x-{1\over 2}}
\ee
\be\label{s5}
S_b\left({1\over b}+bx\right)\to {2\pi(2\pi b^2)^{x-{1\over 2}}\over \Gamma(1-x)}
\ee
\be\label{g1}
\Gamma_b(bx)\to (2\pi b^3)^{{1\over 2}(x-{1\over 2})}\Gamma(x)
\ee
\be\label{g2}
\Gamma_b(Q-bx)\to \sqrt{2\pi}(2\pi b)^{{1\over 2}({1\over 2}-x)}
\ee

\section{Meijer G-function}
The  Meijer G-function can be defined via the integral \cite{ryzhik}:
\bea
 &&G^{m,n}_{p,q}
\left(x\Bigg|\begin{array}{c}
a_1,\ldots,a_p\\
b_1,\ldots,b_q\end{array}\right)=\\ \nonumber
&&{1\over 2\pi i}\int{\prod_{j=1}^m\Gamma(b_j-s)\prod_{j=1}^n\Gamma(1-a_j+s)\over
\prod_{j=m+1}^q\Gamma(1-b_j+s)\prod_{j=n+1}^p\Gamma(a_j-s)}x^sds
\eea
In this paper we will consider the $G^{3,2}_{3,3}$ function. It admits the decomposition \cite{ryzhik}:

\bea\label{g3332}
&& G^{3,2}_{3,3}
 \left(x\Bigg|\begin{array}{c}
a_1,a_2,a_3\\
b_1,b_2,b_3\end{array}\right)=\\ \nonumber
&&{\Gamma(a_1-a_2)\Gamma(1+b_1-a_1)\Gamma(1+b_2-a_1)\Gamma(1+b_3-a_1)\over
\Gamma(1+a_3-a_1)}x^{a_1-1}\\ \nonumber
&&\times{}_3F_2\left(\begin{array}{c}
1+b_1-a_1,1+b_2-a_1,1+b_3-a_1\\
1+a_2-a_1,1+a_3-a_1;\, x^{-1} \end{array}\right)+\\ \nonumber
&&{\Gamma(a_2-a_1)\Gamma(1+b_1-a_2)\Gamma(1+b_2-a_2)\Gamma(1+b_3-a_2)\over
\Gamma(1+a_3-a_2)}x^{a_2-1}\\ \nonumber
&&\times{}_3F_2\left(\begin{array}{c}
1+b_1-a_2,1+b_2-a_2,1+b_3-a_2\\
1+a_1-a_2,1+a_3-a_2;\, x^{-1}\end{array}\right)
\eea
Here ${}_3F_2$ is the generalized hypergeometric function:
\bea
 {}_3F_2\left(\begin{array}{c}
a, b,c;\\ \nonumber
 d, e: x\end{array}\right)=\sum_{n=0}^{\infty}{(a)_n(b)_n(c)_n\over (d)_n(e)_n}{x^n\over n!}
\eea

where
\be
(a)_n={\Gamma(a+n)\over \Gamma(a)}
\ee
is the Pochhammer symbol.
We will need also the following property of the  Meijer G-function:
\be\label{gxi}
x^{\xi}G^{3,2}_{3,3}
 \left(x\Bigg|\begin{array}{c}
a_1,a_2,a_3\\
b_1,b_2,b_3\end{array}\right)=G^{3,2}_{3,3}
 \left(x\Bigg|\begin{array}{c}
a_1+\xi,a_2+\xi,a_3+\xi\\
b_1+\xi,b_2+\xi,b_3+\xi\end{array}\right)
\ee


\section{${}_3F_2$ and ${}_2F_1$ hypergeometric functions with unit argument}
The ${}_3F_2$ function with the unit argument satisfies
 the identities \cite{slater,wilfrid,whipple}
\bea\label{f321}
 {}_3F_2\left(\begin{array}{c}
a, b,c;\\ \nonumber
 d, e: 1\end{array}\right)=
{\Gamma(1-a)\Gamma(d)\Gamma(e)\Gamma(c-b)\over \Gamma(e-b)\Gamma(d-b)\Gamma(1+b-a)\Gamma(c)}
 {}_3F_2\left(\begin{array}{c}
b, 1+b-d,1+b-e;\\ \nonumber
 1+b-c, 1+b-a: 1\end{array}\right)\\
 +{\Gamma(1-a)\Gamma(d)\Gamma(e)\Gamma(b-c)\over \Gamma(e-c)\Gamma(d-c)\Gamma(1+c-a)\Gamma(b)}
 {}_3F_2\left(\begin{array}{c}
c, 1+c-e,1+c-d;\\
 1+c-b, 1+c-a: 1\end{array}\right)\quad
 \eea

  \bea\label{f322}
 {}_3F_2\left(\begin{array}{c}
a, b,c;\\
 d, e: 1\end{array}\right)=
{\Gamma(d)\Gamma(d+e-a-b-c)\over \Gamma(d-a)\Gamma(d+e-b-c)}
 {}_3F_2\left(\begin{array}{c}
e-c, e-b,a;\\
 d+e-b-c, e: 1\end{array}\right)
 \eea

 Note that if one of the ``upper" arguments of the ${}_3F_2$ function coincide with one of the ``lower" argument it reduces
 to ${}_2F_1$ function:
 \bea
 {}_2F_1\left(\begin{array}{c}
a, b;\\ \nonumber
 c: x\end{array}\right)=\sum_{n=0}^{\infty}{(a)_n(b)_n\over (c)_n}{x^n\over n!}
\eea
${}_2F_1$ function with unit argument is equal to:
 \bea\label{f12}
{}_2F_1\left(\begin{array}{c}
a, b;\\
 c: 1\end{array}\right)={\Gamma(c)\Gamma(c-a-b)\over \Gamma(c-a)\Gamma(c-b)}
\eea

\end{document}